# Electronic structure and photoluminescence properties of Zn-ion implanted silica glass before and after thermal annealing

D.A. Zatsepin[1,2], A.F. Zatsepin[2], D.W. Boukhvalov[3,4], E.Z. Kurmaev[1,2], Z.V. Pchelkina[1,4],

N.V. Gavrilov[5]

[1]*M.N. Miheev Institute of Metal Physics of Ural Branch of Russian Academy of Sciences, 18 Kovalevskoj Str., 620990 Yekaterinburg, Russia*
[2]*Institute of Physics and Technology, Ural Federal University, Mira Str. 19, 620002 Yekaterinburg, Russia*
[3]*Department of Chemistry, Hanyang University, 17 Haengdang-dong, Seongdong-gu, Seoul 133-791, Korea*
[4]*Theoretical Physics and Applied Mathematics Department, Ural Federal University, Mira Street 19, 620002 Yekaterinburg, Russia*
[5]*Institute of Electrophysics, Russian Academy of Sciences-Ural Division, 620016 Yekaterinburg, Russia*

*The results of XPS core-level and valence band measurements, photoluminescence spectra of a-$SiO_2$ implanted by Zn-ions (E=30 keV, D=1·$10^{17}$ $cm^{-2}$) and Density Functional Theory calculations of electronic structure as well as formation energies of structural defects in silica glass induced by Zn-ion implantation are presented. Both theory and experiment show that it is energetically more favorable for implanted zinc ions to occupy the interstitial positions instead of cation substitution. As a result, the Zn-ions embedded to interstitials, form chemical bonds with the surrounding oxygen atoms, formation ZnO-like nanoparticles and oxygen-deficient $SiO_x$ matrix. The subsequent thermal annealing at 900 $^0$C (1 hr) strongly reduces the amount of ZnO nanoparticles and induces the formation of α-$Zn_2SiO_4$ phase which markedly enhances the green emission.*

# 1. Introduction

Embedding metal and semiconducting particles into $SiO_2$ host-matrix has been receiving considerable attention of materials scientists because it is a powerful method to re-build the electronic structure and physical properties of this practically significant wide-gap transparent insulator. Technological application of $SiO_2$ (both in crystalline and amorphous phases) actually is not limited with the use of only stoichiometric forms of silicon dioxide – also doped $SiO_x$ (where $x < 2$) polymorphs are employed for passivation coatings and interlayers in microelectronics [1-2], low-index mid-infrared protecting coatings for mirrors [3], etc. Moreover, exactly embedded metal particles (MP) induce the nanostructuring of the silicon oxide and the major question arising herewith is linked with the final chemical state (formal valency) of MP's which will be fabricated by technological end-process treatment [4].

The most interest during last decades was focused on embedding Zn-metal particles into $SiO_2$ by means of Zn-ion implantation (see e.g. [4-8]). It is not surprising, because the understanding of Zn-ion incorporation process is the point for controlled modification of key electronic properties for $SiO_2$:Zn system that might be employed in the fields of photovoltaics, light-emitting/laser applications, optoelectronics, etc. An accumulated data reported previously [2-8] allow to conclude that most of technologists are using thermal annealing in the range of temperatures from 600 $^o$C up to 700 $^o$C after Zn-ion embedding process in order to obtain the defect-free high-quality Zn-doped $SiO_2$. But these recent results seem far away from being perfect – i.e. Zn-implantation with 160 keV and $1.0 \times 10^{17}$ cm$^{-2}$ with the following 700 $^o$C annealing did not display even the signs of Zn-incorporation into $SiO_2$-host [9]. The situation becomes better when 60 keV Zn-implantation of the same doze and 600 $^o$C annealing had been applied, but in this case Zn-atoms are distributed non-uniformly and in the near surface region of $SiO_2$ substrate with partial oxidation of Zn-metal [5]. Also in the most of cited above papers the authors are concluding about significant transportation of Zn-atoms inside the volume of ion-beam treated host-matrix which is strongly impeding the high-quality uniformly nanostructured $SiO_2$:Zn fabrication.

In the present paper we have studied the electronic structure and luminescence properties of Zn-ion implanted silica glasses ($E = 30$ keV, $D = 1 \times 10^3$ cm$^{-2}$) before and after thermal annealing at 900 $^{\circ}$C (1 hr). The XPS measurements (core-levels and valence bands) are compared with performed density functional theory (DFT) calculations of the electronic density of states (DOS) and the formation energies for different configuration of structural defects that were induced by ion-implantation.

## 2. Experimental and Calculations Details

Silica glass samples of KU-type were implanted with $Zn^+$-ions having 30 keV energy with fluence of $1 \times 10^{17}$ cm$^{-2}$. The KU-type of the glass is a technological signature and means the high-purity optical silica glass of type III, obtained by hydrolysis technology from silicon tetrachloride vapour in oxygen-hydrogen flame. It has initially a high homogeneity, very low concentrations of metal impurities and high content of hydroxyl groups (the so-called "wet" silica glass). All these features provide a high radiation-optical stability and a high transparency in the UV and visible regions. Ion irradiation was performed employing the pulsed mode with pulse duration of 0.4 ms and frequency of 25 Hz. The current density of the beam during the pulse was not more than 0.6 mA/cm$^2$. Thermal annealing of the implanted samples was made at 900 $^{\circ}$C in oxygen media within 1 hour.

X-ray photoelectron spectroscopy (XPS) measurements were made with a PHI XPS Versaprobe 500 spectrometer (ULVAC-Physical Electronics, USA) allowing to achieve an energy resolution of $\Delta E \leq 0.5$ eV for Al $K\alpha$ radiation (1486.6 eV). As in our previous experiments, the samples under study were kept in $10^{-7}$ Pa vacuum for 24 h prior to measurement and then surface chemical state mapping attestation was made. Only samples whose surfaces were free from micro impurities were measured and reported herein. The XPS spectra were recorded using monochromatized Al $K\alpha$ X-ray emission with the X-ray spot size of 100 μm in dia. The X-ray power load delivered to the sample was not more than 25 W in order to prevent X-ray stimulated sample decomposing. Under conditions mentioned, the typical XPS signal-to-noise ratios in our experiments were at least not worth than 10000 : 3. An experimental error for binding energy detection for described above conditions was not more than 0.15 eV according to the statement of XPS spectrometer manufacturer. Finally, the spectra were processed using ULVAC-PHI MultiPak Software 9.3 and then the residual background (BG) was removed using the Tougard approach with Doniach-Sunjic line-shape asymmetric admixture [10]. Well known, that most of the provided background models are self-consistent and they are using Doniach-Sunjic type of asymmetrical line-shapes that are acceptable in most common XPS cases. The advantage of retaining asymmetry in XPS data processing usually strongly apparent when a Tougaard BG is used in order to remove the extrinsic contribution to XPS-spectrum of a metal-like or metal-doped materials. Tougaard approach deals with a so-called "three-parameter universal cross-section" and has established values for a number of compounds, including aluminum and its oxides, silicon, silicon dioxide and others [10], so it is a theoretically based choice. After BG-subtraction, the XPS spectra were calibrated using reference energy of 285.0 eV for the carbon 1$s$ core-level. Exactly such a sequence allows performing much better calibration due to previously removed outer contributions to the XPS line-shape.

An X-ray diffraction (XRD) measurements were made using an X'Pert Pro MRD X-ray diffractometer (Panalytical, Holland) under Cu $K\alpha$ radiation with a 1$^o$ anticaster gap and a PIXcel detector having 3.347$^o$ of active length. The XRD patterns were recorded in a Bragg–Brentano parafocusing geometry with a nickel filter using the secondary beam.

Additionally the samples under study were certified with photoluminescence (PL) measurements. The photoluminescence spectra were recorded under selective excitation with 6.5 eV at room temperature using McPherson VUVAS 1000 PL spectrometer (McPherson, USA) with a 30 W deuterium light source. This system guaranteed meets the requirements of deep and vacuum ultraviolet analysis with the energy resolution of 0.2 eV and less than 0.5 % of recorded PL-data distortion.

The electronic structures of $SiO_2$, $Zn_2SiO_4$ and Zn-doped $SiO_2$ were calculated using the tight-binding linear muffin-tin orbital (TB-LMTO) method [11-12] with the von Barth–Hedin local exchange-correlation potential [13]. The lattice constants and atomic positions corresponding to the $P3_121$ symmetry group of α-quartz were taken from Ref. [14]. The muffin-tin sphere radii were chosen to be $R$(Si) = 1.94 a.u., $R$(O) = 1.6 a.u. and 144 k points in the full Brillouin zone were employed for calculations. The simulation of the Zn defects (Zn(I) and Zn(S)) in the interstitial and in Si sites of α-quartz was performed with the super cell of 24 Si-atoms. The crystal structure data for the tetragonal phase of $Zn_2SiO_4$ with the space group $I42d$ was reproduced from Ref.[15]. Here the muffin-tin sphere radii were chosen to be $R$(Zn) = 2.21 a.u., $R$(Si) = 1.96 a.u. and R(O) = 1.58 a.u., applying 512 k points in the full Brillouin zone for calculations. All the calculated DOSes are presented in Results and Discussion sections.

Density functional theory (DFT) was also used for calculation of formation energies for different configurations of structural defects induced by Zn-ion implantation of $a$-$SiO_2$. These calculations were performed with using of the SIESTA pseudopotential code [16-19], a technique that has been recently successful in related studies of impurities in $SiO_2$ [18]. All calculations were made employing the Perdew–Burke–Ernzerhof variant of the generalized gradient approximation (GGA-PBE) [19] for the exchange-correlation potential. All calculations were made in spin-polarized mode. A full optimization of the atomic positions was carried out during which the electronic ground state was consistently found using norm-conserving pseudopotentials for the cores and a double-ξ plus polarization basis of localized orbitals for Si, Zn and O. The forces and total energies were optimized with accuracies of 0.04 eV Å$^{-1}$ and 1.0 meV, respectively. Calculations of formation energies ($E_{form}$) were performed by considering the supercell both with and without a given defect [18]. For the current modeling of zinc impurity interactions with quartz-like matrix we used the $Si_{24}O_{48}$ supercell (see Fig. 1).

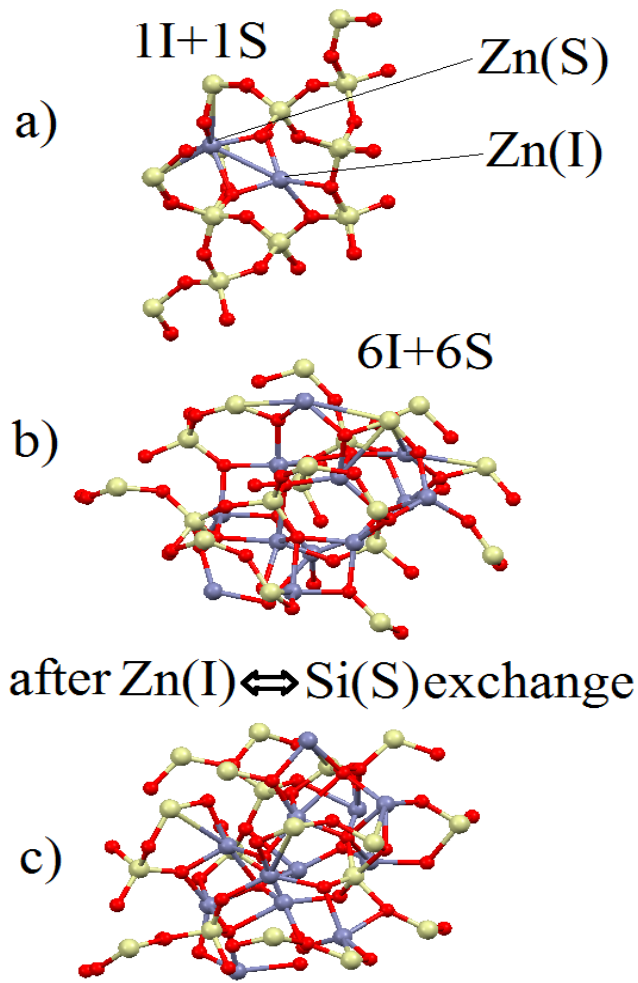

**Figure 1.** (a) An optimized atomic structure of $SiO_2$ cluster at the first stage of ZnO-phase formation; (b) The very beginning of a ZnO-like cluster transformation to $Zn_2SiO_4$; (c) An exchange of interstitial Zn-impurity with Si from $SiO_2$ environment of ZnO-like cluster.

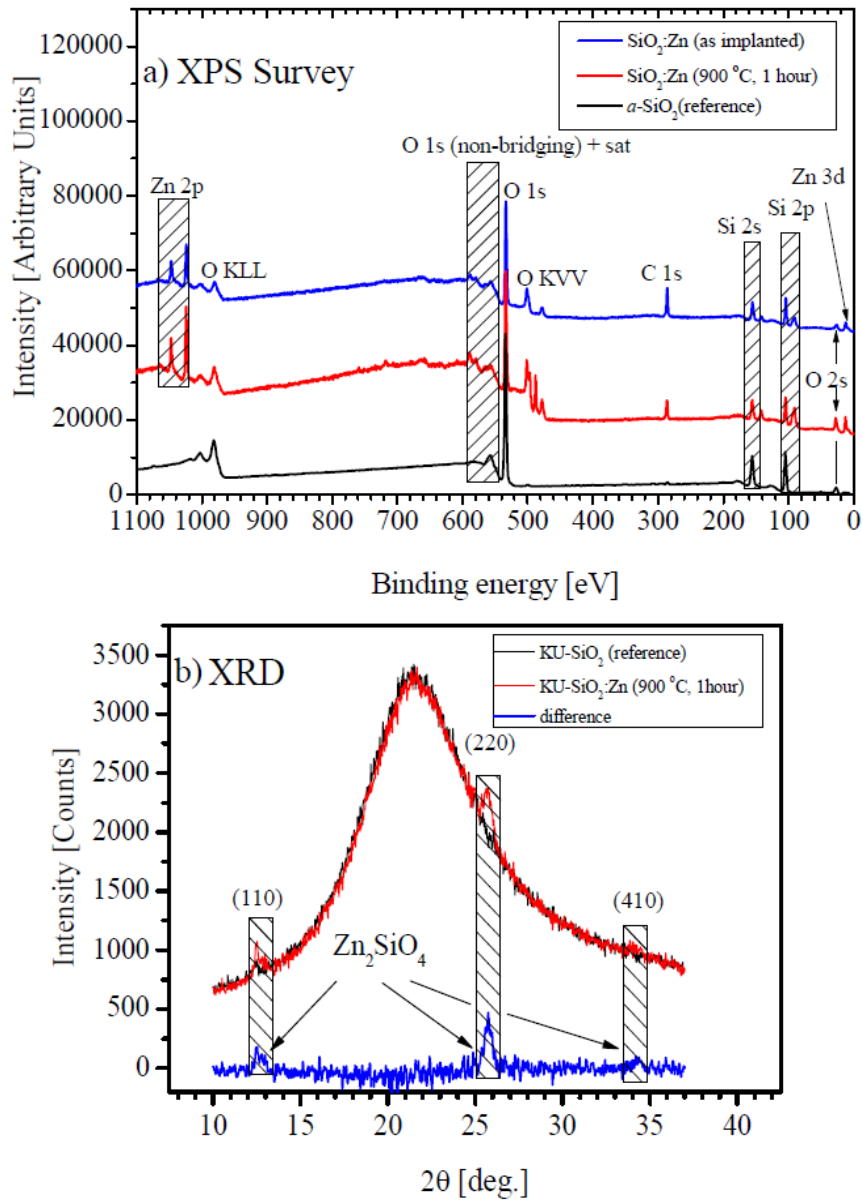

**Figure 2.** (a) XPS Survey spectra of Zn-implanted *a*-SiO$_2$ samples comparing with that for *a*-SiO$_2$ XPS external standard (reference); (b) The results of X-ray diffraction (XRD) measurements for thermally annealed Zn-implanted *a*-SiO$_2$.

## 3. Results and Discussion

X-ray photoelectron elemental analysis was performed on the basis of external XPS standard for the samples under study and is presented at Fig. 2a. These XPS-data within used spatial (100 μm) and depth (~ 3–5 nm) resolution show that samples under investigation do not contain any additional impurities except Zn-dopant that has been embedded into *a*-SiO$_2$ host-matrix by means of pulsed ion-doping. Additionally, XRD-data recorded (see Fig. 2b) clearly demonstrate the new peaks, arising in XRD spectrum for thermally annealed Zn-ion implanted silica glass in contrast with untreated *a*-SiO$_2$ host (reference sample), and these peaks exactly correspond to Zn$_2$SiO$_4$-phase [20]. Appearance of additional peaks in XRD spectra which might be interpreted as signature of Zn$_2$SiO$_4$-phase is the key for the further reliable interpretation of the results obtained by XPS and PL measurements.

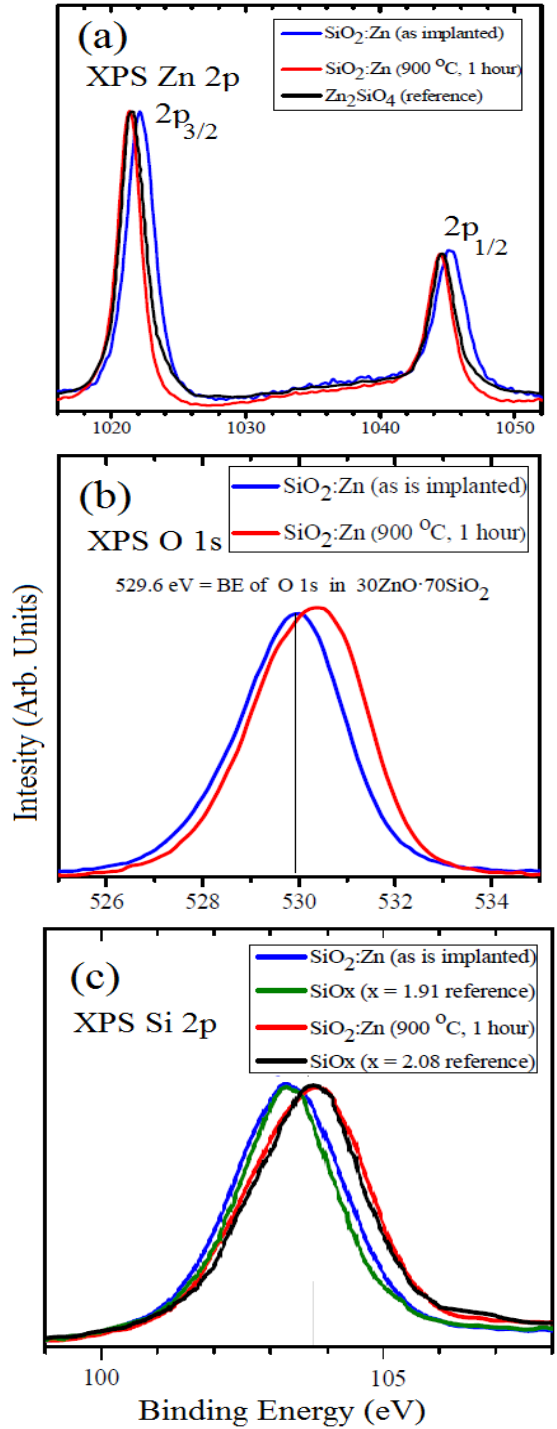

**Figure 3.** (a) XPS Zn 2$p$ core-levels of Zn-implanted $a$-SiO$_2$ and Zn$_2$SiO$_4$ XPS external standard (reference); (b) XPS O 1$s$ core-levels of as is Zn-implanted $a$-SiO$_2$ and after thermal treatment at 900 $^{\circ}$C (1 hr); (c) XPS Si 2$p$ core-levels of $a$-SiO$_2$ after as is implantation and thermal treatment (900 $^{\circ}$C, 1 hr). The appropriate spectra of reference SiO$_x$ with calibrated oxygen nonstoichiometry (x = 1.91 – oxygen deficit and x = 2.08 – oxygen excess) are also shown.

The results of XPS measurements is the evidence that Zn 2*p* core-level of as implanted and annealed samples are shifted relative to each other by 0.9 eV (Figure 3a), wherein the value of binding energy shift between their main maxima is almost twice time larger comparing with XPS measurements error. The binding energy (BE) position of XPS Zn 2$p_{3/2}$ peak for as implanted SiO$_2$:Zn sample well coincides with that of reference bulk ZnO [21]: BE of Zn 2$p_{3/2}$ = 1022.2 eV (as implanted SiO$_2$:Zn) and BE of Zn 2$p_{3/2}$ = 1022.3 eV (ZnO). Further thermal treatment of the implanted sample at 900 $^o$C in the oxygen media within 1 hr results in decreasing of Zn 2*p* core-level binding energy down to 1021.4 eV. Moreover, XPS spectrum line-shape for thermally annealed sample is nearly identical to that for reference Zn$_2$SiO$_4$ [22] (see Fig. 3 (a)) as well as the BE values of Zn 2$p_{3/2-1/2}$ peaks.

Measurements of XPS O 1*s* core-level spectra of the samples under study also demonstrate similar changes of the line-shape and BE values (see Fig. 3b). We also would like to note the fact that O 1*s* energy position for as is implanted sample well coincides with BE O 1*s* for dual oxide 30ZnO·70SiO$_2$ system [23]: BE O 1*s* = 529.8 eV (SiO$_2$:Zn) and BE O 1*s* = 529.6 eV (30ZnO·70SiO$_2$). The difference of 0.2 eV between measured and previously reported in the literature data [23] is, from the one hand, formally nearly close to experimental error but, from the other hand, might be due to the fact that the ZnO–SiO$_x$ (x < 2) composite is forming. This result might be explained by widely known side-effect of ion-implantation — the transport of oxygen within the host-volume [23-24], and as a sequence the oxygen sublattice will become imperfect due to fabricated nonstoichiometry of host-matrix [24-25] (in our case it might be the formation of $SiO_x$ clusters). Also an oxidation of embedded dopant possibly takes place due to the mentioned reasons. Based on discussion above, we can propose that the process of ion-implantation with Zn, but without thermal treatment (as is implanted sample), provides the formation of $ZnO–SiO_x$ composite according to following reaction:

$$SiO_2 + Zn\text{-implantation} \rightarrow ZnO + SiO_x \qquad (1)$$

whereas the thermal annealing of Zn-implanted sample in the oxygen media is forming $Zn_2SiO_4$:

$$SiO_2 + Zn\text{-implantation} \rightarrow ZnO + SiO_x \rightarrow(\text{annealing at } 900\ ^oC, 1\ hr)\rightarrow Zn_2SiO_4 \qquad (2)$$

The conclusion about thermo-stimulated sintering of $Zn_2SiO_4$ oxide after Zn-implantation of $SiO_2$ host well agrees with our XPS data discussed above.

In order to prove or deny the proposed model of formation of nonstoichiometric $SiO_x$ clusters, we performed an additional XPS analysis for Si 2*p* core-levels of as is Zn-implanted samples, implanted and annealed samples, and calibrated external XPS $SiO_x$ standards with a dissimilar content of oxygen (Fig. 3 (b)). XPS spectra given in this figure demonstrate that Si 2*p* core-level spectra of $SiO_x$ (x = 1.91) reference and as implanted $SiO_2$:Zn are identical. This coincidence cannot be random and it is supporting our supposition about implantation-fabricated dual-oxide $ZnO–SiO_x$ system within the volume of ion-modified host matrix according to reaction (1). On the other hand, the appropriate spectra for $SiO_x$ (x = 2.08) reference and implanted $SiO_2$:Zn after annealing in oxygen media are also identical which is in accordance with formation of $Zn_2SiO_4$ phase (reaction 2). This conclusion is additionally supported by XRD measurements (Fig. 2 (b)) which clearly show that additional peaks are appeared for thermally annealed Zn-ion implanted silica glass and they are exactly correspond to $Zn_2SiO_4$-phase [20].

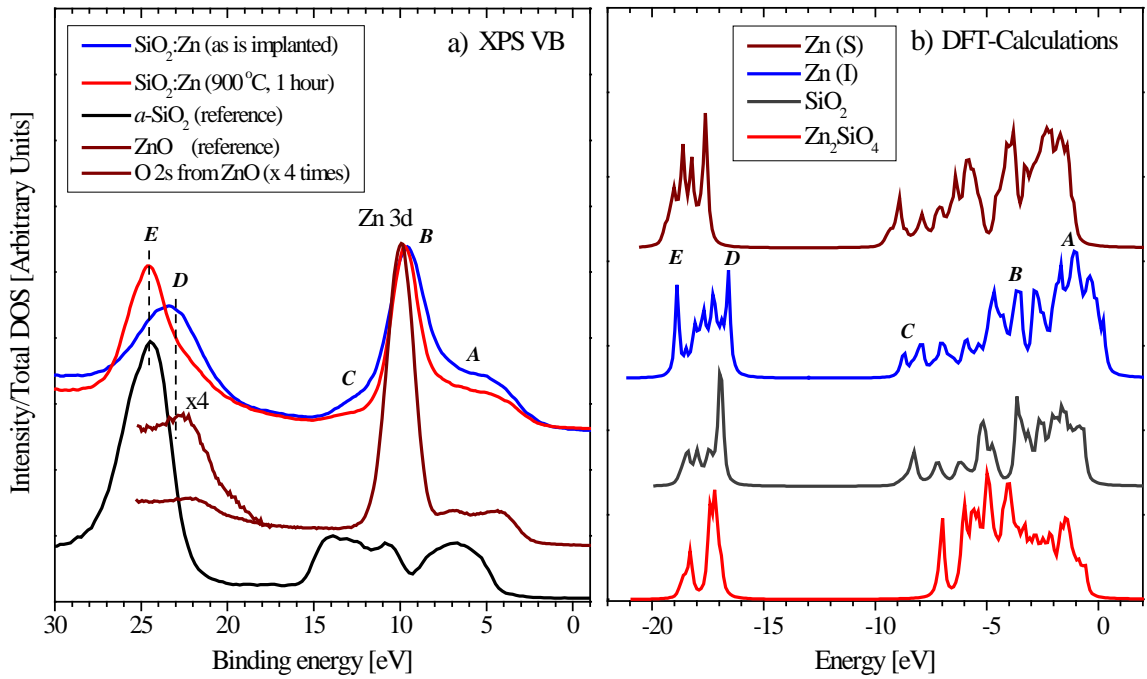

**Figure 4.** (a) XPS VB spectra of as implanted and annealed $SiO_2$:Zn compared with those of ZnO and $a$-$SiO_2$ reference samples; (b) Calculated total DOSes of $Zn_2SiO_4$ and Zn-doped $SiO_2$.

Comparison of XPS VB spectra (Fig. 4a) of ion-modified $a$-$SiO_2$ and reference samples demonstrates further differences between $SiO_2$:Zn before and after thermal annealing. These differences are arising (*i*) within the energy range of 21–27 eV (*D-E*), where the O 2$s$ states are located and (*ii*) at the bottom (*C*) and top (*A*) of the valence band, where the relative intensity of *C*/*B* and *C*/*A* peaks is lower for implanted and annealed sample in comparance with as is implanted one. One can see that the energy position of O 2$s$-band for as is implanted sample is close to that for ZnO which is in accordance with our conclusion about the formation of ZnO-like nanoparticles in $a$-$SiO_2$:Zn. On the other hand, the energy distribution of O 2$s$-band is wider than that of ZnO and spread to that of $a$-$SiO_2$, because the oxygen-atoms form different bonds with zinc and silicon atoms (as in ZnO and $SiO_x$), meaning that ZnO nanoparticles are formed in $SiO_x$ matrix. In XPS VB of thermally annealed sample the energy position of O 2$s$-band is very close to that of $a$-$SiO_2$, because thermal annealing induces the decreasing of ZnO-like nanoparticles contribution to the overall electronic structure of final composite and formation of $Zn_2SiO_4$ phase with the same tetrahedral $SiO_4$-units (as in ordinary silica glass) occurs. These conclusions are also supported by DFT calculations of the electronic structures of $SiO_2$-host, $Zn_2SiO_4$ and Zn-doped $SiO_2$ with Zn-ions embedded into the lattice sites as substitution Zn(S) and interstitials Zn(I) (see Fig. 4 (b)). As seen, the O $2s$-bands are rather similar in $SiO_2$ and $Zn_2SiO_4$. On the other hand, the reduction of $A$-subband contribution and the absence of $C$-subband is observed in the calculated total DOSes of $Zn_2SiO_4$. This is well coincides with XPS VBs spectra of thermally annealed $a$-$SiO_2$:Zn.

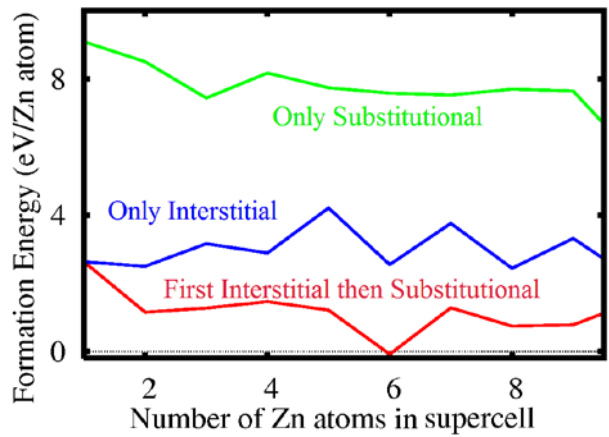

**Figure 5.** Formation energies for the three different scenarios of Zn-incorporation into $a$-$SiO_2$ matrix.

To understand the mechanism of ZnO-like nanoparticles formation in as is implanted $a$-$SiO_2$ and further $Zn_2SiO_4$-phase transition after thermal annealing, we have performed the DFT calculations of formation energies for three possible scenarios of Zn-impurities incorporation into $SiO_2$-matrix: (*i*) the only substitutional impurities (S), (*ii*) the only interstitial defects (I), and, (*iii*) primarily the insertion of interstitial impurity and only then the substitution of Si-atom by Zn (mixed configuration of interstitial and substitutional impurities). According to the calculated formation energies, presented at Fig. 5, the substitution of $Si^{4+}$ atoms by $Zn^{2+}$ cannot be energetically favorable because of the appeared dangling bonds on oxygen atoms, caused by the difference in oxidation state of silicon and zinc ions. This is well illustrated by DFT calculations of total DOSes of Zn-doped $SiO_2$ with Zn-substituted ions Zn(S) which contradict with experimental XPS VB spectra (compare with the data at Fig. 4a). The insertion of Zn-impurities only as interstitial void requires less energy than substitution, but the lowest formation energies for all studied Zn-concentrations as Zn-impurity were found as a combination of interstitial and substitutional impurities. So the initial steps of this process in the studied supercell can be described as:

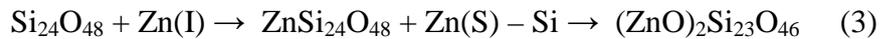
$$Si_{24}O_{48} + Zn(I) \rightarrow ZnSi_{24}O_{48} + Zn(S) - Si \rightarrow (ZnO)_2Si_{23}O_{46} \quad (3)$$

where the first step is an insertion of the interstitial impurity that is over 5 eV energetically favorable with respect to the substitutional defect (Fig. 5). The second step is a substitution of Si-atom in the vicinity of interstitial impurity. This stage is more than 1 eV energetically favorable than formation of the next interstitial defect (Fig. 5). Thus we can conclude that the ZnO-phase formation in Zn-implanted $a$-$SiO_2$ takes place as a combination of interstitial and substitutional Zn-impurities.

Within the next stage of our modeling, we check the energetics of the very beginning of $Zn_2SiO_4$-phase formation. This process was simulated as an exchange of atoms between ZnO-like clusters and $SiO_2$-matrix. We substitute an interstitial Zn-impurity with Si-atom from $SiO_2$ part of supercell in the forward and backward direction and then compare the total energies of the supercells before and after such Zn(I)↔Si(S) exchange. Figure 1c illustrates the transformation of supercell atomic structure after this process. Each step for the process mentioned above requires the energies of 2~3 eV order even without taking into account the migration energies which are less energetically favorable than the formation of ZnO-like clusters. The latter is requiring relatively high-temperatures for the transformation of ZnO-like nanoparticles into $Zn_2SiO_4$ phase. Note, that in this simulation we estimate the energetics only for the first stage of ZnO + SiO$_2$ → Zn$_2$SiO$_4$ transformation, because the simulation of direct phase transition caused by thermal annealing requires the so-called large-scale molecular dynamic calculations. This means that in order to take into account the local oxygen and silicon concentrations in Zn-rich areas, an extremely large supercell have to be used. Unfortunately, such calculations are out of possibilities for the current computational facilities.

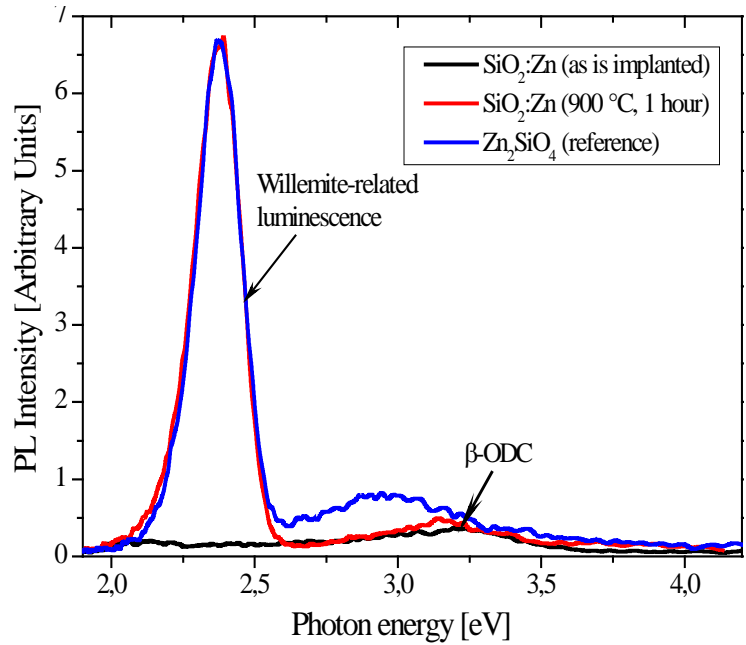

**Figure 6.** Photoluminescence (PL) spectra of *a*-SiO$_2$ host after Zn-implantation, implantation and annealing at 900 °C (1 hr) measured at 300 K.

In order to verify the discussed above our results of XPS certification for the samples under study, their photoluminescence (PL) measurements at 300 K were additionally performed (see Fig. 6). From this figure could be clearly seen that PL spectrum of implanted and annealed sample clearly exhibits the main high-intensity narrow maximum, located at ~ 2.35 eV and the low-intensity symmetrical wide band centered at ~ 3.25 eV. This band is shifted to the lower photon energies in the spectrum of reference α-Zn$_2$SiO$_4$ and is placed at about 2.95 eV, but the main maximum well coincides with that for SiO$_2$:Zn (annealed at 900 °C, 1 hour). At the same time the main maximum is totally absent in the PL spectrum of as is implanted sample (not annealed). The main relatively narrow maximum located at ~ 2.35 eV is arising because of the radiative transitions within the intrinsic luminescence centers. We assume that in this case there are luminescent centers of intrinsic defect type that are present in the lattice of α-$Zn_2SiO_4$ (recall, that it is absent in as is implanted sample). Also we have to note, that the origin of intrinsic luminescence of $Zn_2SiO_4$ as well as theoretical models still remains foggy so this question needs further separate and deep study. At the same time the low-intensity and wide band observed at 3.25 eV is believed to be due to triplet-singlet $T_1 \rightarrow S_0$ transition in oxygen-deficient centers β-ODC's that are present in untreated $SiO_2$-host [26]. From the identity of the main maximum for reference α-$Zn_2SiO_4$ and *a*-$SiO_2$:Zn (annealed at 900 $^oC$, 1 hr) we can conclude with the high-probability that α-$Zn_2SiO_4$ is formed after implantation and annealing. From PL-spectrum of as is implanted sample we can only realize that oxygen deficient centers (ODC's) are appeared in the glassy network of *a*-$SiO_2$, fabricating oxygen nonstoichiometry that had been detected by analyzing the XPS Si 2*p* core-levels (see Fig. 3 c). Finally, the reported PL-based conclusions are not contradicting with discussed above XPS core-level and valence band analysis as well as XRD data.

**Conclusions**

To conclude we have measured XPS core-levels, valence bands, photoluminescence spectra and XRD patterns of Zn-ion implanted silica glass before and after thermal annealing at 900 $^0C$ (1 hour). Our results were compared with performed DFT calculations of electronic structure and formation energies for different configuration of structural defects induced by ion implantation. Basing on comparison of theory and experiment we can conclude that in as implanted silica the stable ZnO nanoparticles are formed whereas the thermal annealing induces the formation of $Zn_2SiO_4$ species with strongly increased green emission. Thereby, the fabrication of luminescing $Zn_2SiO_4$ nanoparticles in Zn-implanted *a*-$SiO_2$ host might be presented as a two-step solid-state process which is including the formation of ZnO nanoparticles as an intermediate stage with the following their interaction with the glassy network of $a$-SiO$_2$ host-matrix.

The observed relatively high-intensive green emission of implanted and thermally treated $a$-SiO$_2$:Zn samples might be linked with the intrinsic defects in the newly fabricated nanoscaled Zn$_2$SiO$_4$-phase. This fact seems to be very important in terms of understanding the physical origin and the peculiarities of optical features of advanced photonic nanomaterials synthesized with the help of ion-beam treatment technologies. Besides, we think that the results obtained and reported in the current paper are of concrete practical significance for the future study of possibility to modify controllably the functional characteristics of new generation optoelectronic and microelectronic devices.

**Acknowledgements**

The preparation of $a$-SiO$_2$ samples, ion-implantation treatment and photoluminescence measurements were supported by Russian Foundation for Basic Research (Projects RFBR Nos. 13-08-00568 and 13-02-91333) and the Government Assignment of the Russian Ministry of Education and Science (3.1016.2014/K).The XPS measurements and DFT calculations were supported by Russian Science Foundation (Project No. 14-22-00004).